- # Highly tunable quantum light from moiré trapped excitons

H. Baek,[1][†][*] M. Brotons-Gisbert,[1][†] Z. X. Koong,[1] A. Campbell,[1] M. Rambach,[1] K. Watanabe,[2] T. Taniguchi,[2] and B. D. Gerardot[1][*]

[1]Institute of Photonics and Quantum Sciences, SUPA, Heriot-Watt University, Edinburgh EH14 4AS, UK
[2]National Institute for Materials Science, Tsukuba, Japan
e-mail address: H.Baek@hw.ac.uk, B.D.Gerardot@hw.ac.uk

Photon antibunching, a hallmark of quantum light, has been observed in the correlations of light from isolated atomic and atomic-like solid-state systems. Two-dimensional semiconductor heterostructures offer a unique method to create a quantum light source: a small lattice mismatch or relative twist in a heterobilayer can create moiré trapping potentials for excitons which are predicted to create arrays of quantum emitters. While signatures of moiré trapped excitons have been observed, their quantum nature has yet to be confirmed. Here we report photon antibunching from single moiré trapped interlayer excitons in a heterobilayer. Via polarization resolved magneto-optical spectroscopy, we demonstrate the discrete anharmonic spectra arise from bound band-edge electron-hole pairs trapped in moiré potentials. Finally, using an out-of-plane electric field, we exploit the large permanent dipole of interlayer excitons to achieve large DC Stark tuning, up to 40 meV, of the quantum emitters. Our results confirm the quantum nature of moiré confined excitons and open opportunities to investigate their inhomogeneity and interactions between the emitters or tune single emitters into resonance with cavity modes or other emitters.

## I. INTRODUCTION

The ability to stack unlimited combinations of atomic layers with arbitrary crystal angle ($\theta$) has opened a new paradigm in quantum material design. For example, easily tunable Bloch minibands emergent in moiré lateral superlattices have enabled remarkable observations with graphene heterostructures, such as nearly flat bands with narrow bandwidths at specific $\theta$ [1] that can lead to superconductivity [2] and correlated insulator states [3]. Beyond graphene, new opportunities arise with transition metal dichalcogenide (TMD) semiconductors, where band-edge electrons and holes located at two degenerate, but inequivalent, corners of the Brillouin zone (±K valleys) form excitons with a strong Coulomb interaction [4]. Due to strong spin-orbit coupling, the carriers exhibit locked spin and valley degrees of freedom. The in-plane 2π/3 rotational ($\hat{C}_3$) symmetry of the monolayer TMD crystal structure generates valley contrasting optical selection rules for strongly bound excitons [5]. Stacking any two different monolayer TMDs creates a heterobilayer with Type II band-alignment [6] which exhibits spatially indirect interlayer excitons (IXs) with highly tunable photoluminescence (PL) [7-10]. The twist of a TMD heterobilayer changes the displacement of the constituent ±K valleys, determining the coupling of IXs to the light-cone [11]. Similar to graphene bilayers (BLs), the constituent TMD monolayers (ML) also interact with each other and create moiré potentials dependent on $\theta$ which can hybridize wavefunctions across both layers [12-14] or lead to uniform high-density arrays of quantum emitters [15, 16] or topological bands whose properties can be manipulated by electric or strain fields [17-19].

Absorption and PL of TMD heterobilayer samples have recently been investigated to probe for moiré trapping potentials [20-24]. In the limit of low temperature and weak excitation, PL spectra exhibit sharp lines, similar to III-V or WSe$_2$ quantum dots [25-27] except for strong helical polarization due to $\hat{C}_3$ symmetry of the constituent crystal lattices and a notable absence of observable fine-structure [20, 21]. In addition, highly uniform g-factors dependent on relative layer twist are observed, clear fingerprints of the spin and valley configurations for excitons composed of band-edge electrons and holes at the ±K points. Finally, the helical polarization appears to be determined by the atomic registry. Combined, these observations provide compelling evidence for moiré trapped IXs. Nevertheless, ambiguity remains about their precise nature. Do the sharp spectral features arise from single trapped excitons? Why are the spectral features inhomogeneous, unlike the g-factors? Here we provide unambiguous proof of the quantum nature of the moiré trapped IXs via the observation of photon antibunching. This opens a route to investigate second-order cross correlations among the distinct spectral peaks and to understand the full nature of moiré quantum emitter arrays. Further, by incorporating the moiré trapped IXs into a device which enables an applied out-of-plane electric field, we achieve 40 meV tuning of the quantum emitters via the DC Stark effect. This enables a precise measurement of the large permanent dipole of the quantum emitters which results from the electron-hole pair separation in the heterobilayer. Ultimately, our results may lead to engineering highly-tunable arrays of coherent quantum emitters and spin-photon interfaces.

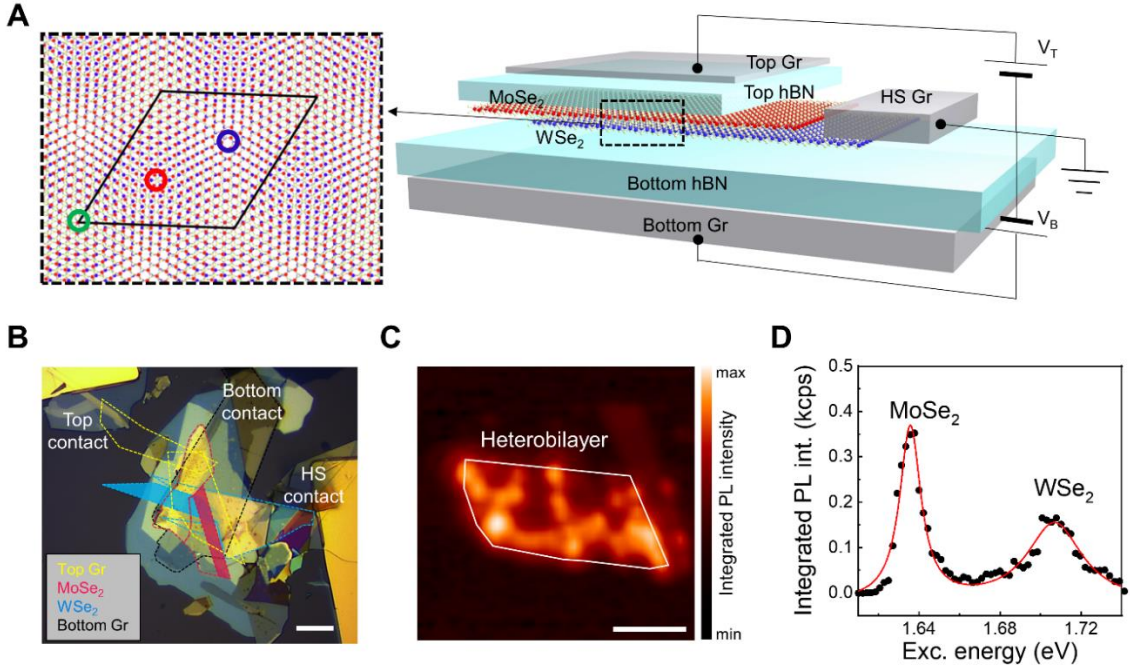

**Fig. 1. MoSe$_2$/WSe$_2$ moiré heterobilayer.** (**A**) Schematic of Sample 1, the moiré heterobilayer with dual gates. Graphite (Gr) layers are used as contacts for top, bottom, and heterostructure (HS), and hBN layers are used for both dielectric layers and encapsulation of heterobilayer. Illustration of a moiré superlattice with a twist angle of ~60° are displayed in dashed box. The moiré supercell is represented as black solid line, and the three circles (in red, blue, and green) indicate the high symmetry points under $\hat{C}_3$ operation. (**B**) Optical micrograph of Sample 1. The top and bottom graphite layers are represented as yellow and black dashed lines respectively, and the MoSe$_2$ and WSe$_2$ flakes are outlined in magenta and blue dashed lines respectively. The filled regions represent ML TMDs. (**C**) Low temperature spatial PL intensity map of the interlayer excitons (energy range of 1.38–1.40 eV). The heterobilayer is outlined with a white solid line. (**D**) Low-temperature PLE intensity plot of a representative IX in the heterostructure, showing two resonances corresponding to the intralayer exciton states in ML MoSe$_2$ and ML WSe$_2$. Scale bars in (B) and (C) are 20 and 5 μm, respectively.

## II. MoSe$_2$/WSe$_2$ MOIRÉ HETEROBILAYER SAMPLES

Our heterobilayer samples consist of ML MoSe$_2$ and ML WSe$_2$ encapsulated by hexagonal boron nitride (hBN). A MoSe$_2$/WSe$_2$ heterobilayer stacked with a small twist angle forms a periodic moiré superlattice as presented in Fig. 1A. To apply vertical electric field to the moiré heterobilayer, we first investigate a dual gated device (Sample 1) fabricated using graphite and hBN as electrical contact and dielectric layers respectively, as shown in Fig. 1A. Figure 1B shows an optical micrograph of the fabricated device. As determined by the cleaved edges of MoSe$_2$ and WSe$_2$ and Landé g-factors described below, the twist angles for the heterobilayer samples we investigate are considered to be close to 60°. Graphite layers, which form the top and bottom gates of heterobilayer, were connected to pre-patterned Au electrodes. The thicknesses of the top and bottom hBN are determined via high-resolution ellipsometry to be 17.4±0.2 nm and 18.2±0.3 nm, respectively. A second moiré heterostructure sample (Sample 2) is engineered for enhanced collection efficiency. Here, we place the heterostructure on a gold mirror and choose the hBN bottom layer thickness (96 nm) to position the heterobilayer at an antinode of the electric field to create a planar antenna [28]. This helps improve the signal of the spatially indirect IXs, which has an intrinsically small oscillator strength [7].

A low temperature (4 K) confocal PL intensity map from Sample 1 in a photon energy range of 1.38–1.40 eV is shown in Fig. 1C. Bright emission from the entire region of heterobilayer is detected, indicating formation of a highly uniform interface between MoSe$_2$ and WSe$_2$ layers. In addition to the heterobilayer, a BL MoSe$_2$ region can be also identified in the top right region of the map by its indirect emission. To confirm the origin of the heterobilayer PL signal arises from IXs, we perform PL excitation (PLE) spectroscopy, in which a continuous wave (CW) excitation laser is scanned from 1.61 eV to 1.75 eV while monitoring the intensity of IX PL at ~1.4 eV. Figure 1D, the PLE spectrum, features two prominent resonances at ~1.63 eV and ~1.70 eV, which correspond to the absorption of the intralayer 1s state of A excitons in ML MoSe$_2$ and WSe$_2$, respectively. To exploit the high quantum yield at resonant excitation of MoSe$_2$, most of the PL spectroscopy results reported here are performed with an excitation energy of 1.63 eV. Similar results are obtained from Sample 2.

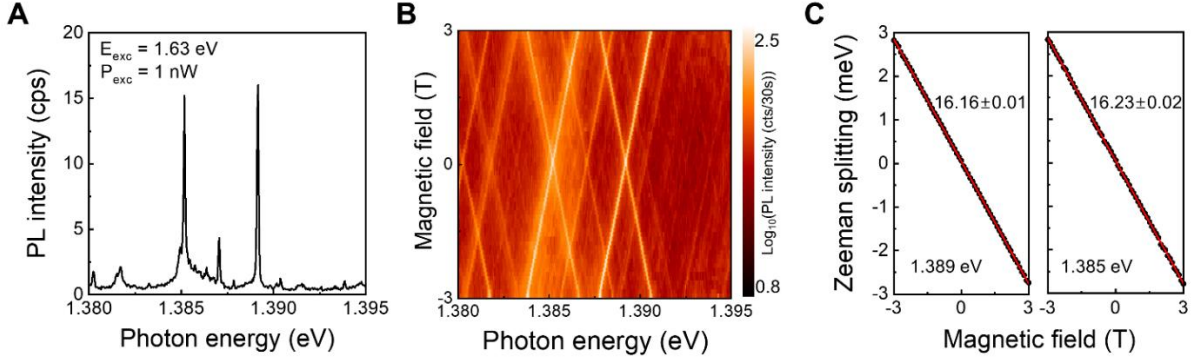

**Fig. 2. Magneto-PL spectroscopy of moiré trapped excitons.** (**A**) A representative PL spectra of moiré IXs with an excitation energy of 1.63 eV at 1 nW power. (**B**) Magnetic field dependence of the moiré IXs in the 2*H*-type MoSe$_2$/WSe$_2$ heterostructure as a function of the applied out-of-plane magnetic field. (**C**) Plots of Zeeman splitting energy versus magnetic field with linear fits for the peaks at 1.389 and 1.385 eV.

## III. MAGNETO-PL SPECTROSCOPY

Figure 2A shows a representative low temperature confocal PL spectrum of moiré IXs. The PL spectrum reveals several discrete spectra with emission energies at ~1.39 eV, in agreement with recently reported values [20, 21]. Peak linewidths of ~100 μeV are observed at low excitation powers; the peak at 1.389 eV has a linewidth of 80 μeV. Such linewidths are similar to previously reported values in moiré trapped IXs and two orders of magnitude smaller than ones from broad IXs (typically 7–30 meV) [8-10, 29, 30]. To confirm the discrete peaks arise from band-edge states at the ±K points, we perform magneto-optical spectroscopy in Faraday configuration. A clear linear Zeeman splitting for each IX peak is observed, as shown in Fig. 2B and C, and an average g-factor of –16.2 is extracted, in strong agreement with the gyromagnetic ratios reported for IX ensemble emission [31] and moiré trapped IXs with 2*H* stacking [20, 21]. Notably, the linear Zeeman splitting persists to very small applied magnetic fields, indicating a lack of fine-structure splitting which can arise due to an asymmetric confinement potential. Additionally, polarization-resolved PL reveals the moiré trapped IX exhibit circular polarization which is co-polarized to the excitation polarization (for excitation resonant with the 1s intralayer resonance of WSe$_2$) (see Supplementary Fig. S1 and S2). The combination of g-factor values and co-polarized emission implies that moiré excitons are confined in the same atomic configuration, likely $H_h^h$ [15, 21], indicated as blue circle in Fig. 1A. Here *H* corresponds to $\theta$ ~60° while the superscript and subscript represent the hexagon center of the electron and hole layers, respectively. Together, these results are compelling evidence that the IXs are composed of bound band-edge electron-hole pairs at the ±K points trapped in rotationally symmetric moiré potentials.

## IV. DC STARK TUNING

Using the top and bottom gates, the dependence of moiré IXs on external electric field is investigated as shown in Fig. 3A and B. Since the thicknesses of the top and bottom hBN are very similar, gate voltages for top ($V_T$) and bottom ($V_B$) are set to the same magnitude but opposite direction to apply vertical electric field without strongly affecting the Fermi energy of the heterobilayer. The peak positions of all of moiré IXs linearly shift with applied electric field. Three representative peaks, indicated as E1, E2 and E3, are highlighted in Fig. 3A and B. A tuning range of ~40 meV is observed, much larger than that of monolayer WSe$_2$ quantum dots which have minimal out-of-plane permanent dipole [27, 32]. Figure 3B shows a plot of photon energy versus gate voltage for E1, E2 and E3 peaks. Using the equation $\Delta U = -pE$ (where $\Delta U$ is the linear Stark shift, $p$ is the out-of-plane electric dipole moment, and $E$ is the vertical electric field) with $E = \frac{(V_T - V_B)}{t_{hBN}} \frac{\varepsilon_{TMD}}{\varepsilon_{hBN}}$ (where $\varepsilon_{TMD} = 7.2$ and $\varepsilon_{hBN} = 3.8$ are the relative permittivity of TMD and hBN, respectively [33, 34], $t_{hBM}$ is the total thickness of top and bottom hBN), the average electrical dipole moment is calculated as 429±4 meVnmV$^{-1}$. Since the magnitude of dipole moment is expressed as $p = ed$, where $e$ is the single electron charge and $d$ is the separation of electron-hole, $d$ is estimated as 0.429±0.004 nm.

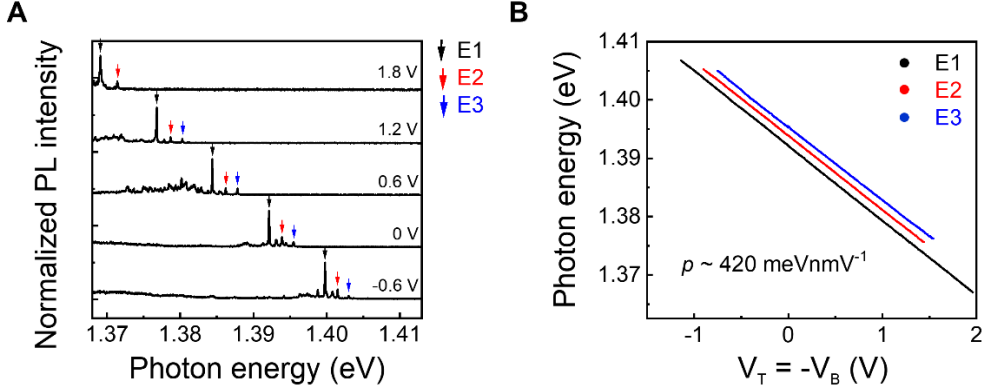

**Fig. 3. Stark tuning of moiré trapped excitons.** (**A**) PL spectra of IXs at different gate voltages. Three representative peaks are indicated as E1, E2, and E3. (**B**) Plot of emission energy versus gate voltage for E1, E2 and E3 peaks. Values of electrical dipole moments of ~ 420 meVnmV$^{-1}$ are determined by the linear fits.

## V. PHOTON ANTIBUNCHING

To characterize the quantum nature of moiré IXs, the power-dependent PL and second-order correlation function $g^{(2)}(\tau)$ are measured for a single emitter. These measurements are performed on Sample 2 which is engineered for enhanced collection efficiency. We identify Sample 2's MoSe$_2$/WSe$_2$ stacking orientation is ~60° by the IX spin-valley fingerprints revealed in the magneto-optics (see Supplementary Fig. S3). Figure 4A shows a PL spectrum which includes the emitter at 1.401 eV which we target for the photon antibunching experiment, chosen due to its relative brightness and minimal background. The emission intensity of the moiré IX saturates with increasing excitation power, as shown in Fig. 4B, hinting at the confined nature. The fit is based on $I = I_{sat}(\frac{P_{exc}}{P_{exc}+P_N})$, where $I$ is the PL intensity, $I_{sat}$ is the saturation intensity, $P_{exc}$ is the excitation power, and $P_N$ (0.47 µW) is the excitation power at which $I = I_{sat}/2$. The emission peak is spectrally filtered for time-resolved PL (TRPL) and $g^{(2)}(\tau)$ measurements. Figure 4C shows the TRPL intensity trace of the emitter. The solid red line represents a fit with a single exponential decay function with a decay time of $T_1 = 12.1±0.3$ ns, within the range reported for IX ensemble emission in MoSe$_2$/WSe$_2$ heterobilayers (2–100 ns) [7, 31, 35]. Next, we send the filtered spectrum to a Hanbury Brown-Twiss interferometer. Figure 4D shows the second-order photon correlation statistics. The red solid line represents a fit using $g^{(2)}(\tau) = 1 - \rho^2 e^{-|\tau|/\tau_c}$, where $\tau$ is the time delay between two consecutive detected photons, $\tau_c$ is the decay time, and $\rho=SBR/(SBR+1)$, with $SBR$ the signal-to-background ratio. The fit reveals $g^{(2)}(0) = 0.28 ± 0.03$ and $\tau_c = 4.3 ± 0.2$ ns. The orange shadowed area represents the Poissonian interval error associated with the experimental determination of $g^{(2)}(\tau)$. The $g^{(2)}(0)$ value is well below the threshold of 0.5, unambiguously proving the quantum nature of the light emitted by the moiré trapped IXs. Finally, the black dashed line and the grey shadowed area represent the average and error interval of the experimental limitation for $g^{(2)}(0)$, respectively, due to non-filtered emission background (Supplementary Section S3) which results in an average $SBR$ of 6.4. The results suggest that spectrally isolated moiré trapped IX offer potential as high-purity single-photon sources and coherent spin-photon interfaces.

## VI. DISCUSSION

Via the observation of photon antibunching, we demonstrate the quantum nature of discrete anharmonic spectra from moiré confined excitons in a two-dimensional heterostructure. The quantum emitters observed from the MoSe$_2$/WSe$_2$ heterobilayers are identified to originate from confined IXs by a moiré potential as confirmed by magneto-optical spectroscopy. The uniform g-factor, lack of fine-structure splitting, and helical polarization are distinguishing features of the moiré potential, which preserves the intrinsic $\hat{C}_3$ symmetry of the constituent crystal lattices. In contrast, localized excitons in monolayer TMDs generated by extrinsic defects or strain exhibit a large fine-structure splitting, large variations in g-factor and linearly polarized emission [26, 27].

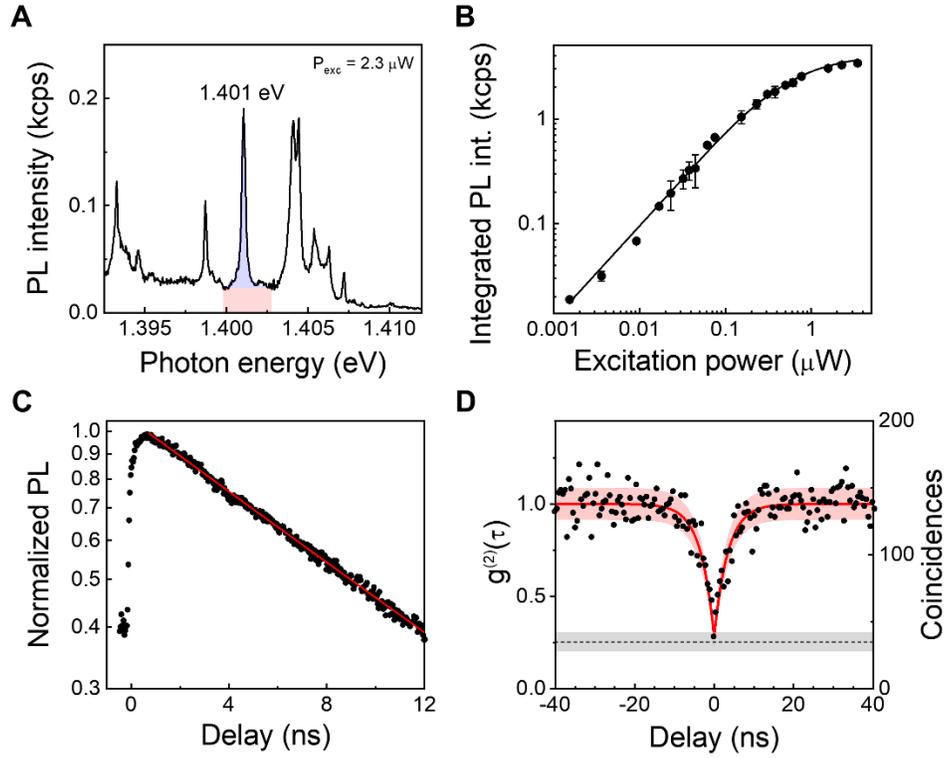

**Fig. 4. Quantum nature of moiré IXs.** (**A**) PL spectrum from Sample 2. The emitter at 1.401 eV is used for power-dependence, TRPL, and second-order photon correlation measurement. The blue and red regions represent the estimated PL signal from the emitter and the background. (**B**) Integrated PL intensity of a single-emitter at different excitation powers. (**C**) Time-resolved normalized PL intensity of the single-emitter under 80 MHz pulsed excitation at 1.63 eV with an average excitation power of 4 μW. Red solid line shows a single exponential decay fit to the experimental data, revealing a lifetime of 12.1±0.3 ns. (**D**) Second-order photon correlation statistics using 2.3 μW CW excitation at 760 nm (black dots) show clear antibunching. The red solid line represents a fit of the experimental data, revealing a $g^{(2)}(0) = 0.28 \pm 0.03$. The red shadowed area represents the Poissonian interval error associated to the experimental determination of $g^{(2)}(\tau)$. The black dashed line represents the experimental limitation for $g^{(2)}(0)$ due to non-filtered emission background. The grey shadowed area shows the error interval in the determination of the limitation for $g^{(2)}(0)$.

Additionally, we establish that the emission energy of moiré IXs can be highly and reliably tuned using the DC Stark effect: we achieve 40 meV tuning in total. Since electrons and holes reside in spatially separated layers, the out-of-plane electrical dipole moment of IXs is much larger than that of intralayer excitons. The electron-hole separation, $d$, is estimated as 0.429±0.004 nm, in rough agreement with previously estimated values (0.5–0.6 nm) of broad interlayer excitons [9, 10]. The slightly smaller value for $d$ obtained here might be due to the interaction between the two layers which causes the moiré potential and provides the carrier localization. Moiré IXs are confined in a specific atomic registry, and it is expected that the interlayer distances at trapping sites are 0.6–2 Å [15, 36] closer than other locations. Thus, moiré trapped IXs can have slightly reduced values of electric dipole moment compared to broad (non-trapped) IXs where the modulation of interlayer distance might be averaged out. The high tunability of moiré IXs combined with the single photon nature enables the pursuit of promising technological applications. The energy tuning can be used to bring moiré quantum emitters into resonance with a cavity for Purcell enhancement for the generation of indistinguishable photons.

From a fundamental point of view, the unambiguous demonstration of quantum light emission from moiré confined IXs is a significant building block to better understand the underlying physics of excitons in moiré superlattice potentials. Second-order cross correlation measurements between neighboring spectral peaks can be used to understand their relationship, addressing questions regarding the perceived inhomogeneity in the moiré superlattice and the possibility of strong interactions within the excitonic superlattice which can lead to superradiance, topological moiré minibands, or the realization of a tunable Mott-Hubbard Hamiltonian [17-19, 37-41].

# ACKNOWLEDGEMENTS

**Funding:** The EPSRC (EP/P029892/1; EP/S000550/1); the ERC (No. 725920); EU Horizon 2020 research and innovation program (No. 820423); Elemental Strategy Initiative conducted by the MEXT, Japan; and CREST (JPMJCR15F3). B.D.G. is supported by a Wolfson Merit Award from the Royal Society and a Chair in Emerging Technology from the Royal Academy of Engineering.

**Author contributions:** B.D.G. conceived and supervised the project. H.B. fabricated the samples. K.W. and T.T. supplied the hBN crystal. H.B. and M.B.-G. performed the experiment assisted by Z.X.K., A.C., and M.R.. H.B. and M.B.-G. analyzed the data, assisted by B.D.G.. H.B., M.B.-G., and B.D.G. cowrote the paper with input from all authors. H.B. and M.B.-G. contributed equally to this work.

[1] R. Bistritzer, A. H. MacDonald, Moiré bands in twisted double-layer graphene. *Proc. Natl. Acad. Sci. U. S. A.* **108**, 12233–12237 (2011).

[2] Y. Cao, V. Fatemi, S. Fang, K. Watanabe, T. Taniguchi, E. Kaxiras, P. Jarillo-Herrero, Unconventional superconductivity in magic-angle graphene superlattices. *Nature* **556**, 43–50 (2018).

[3] Y. Cao, V. Fatemi, A. Demir, S. Fang, S. L. Tomarken, J. Y. Luo, J. D. Sanchez-Yamagishi, K. Watanabe, T. Taniguchi, E. Kaxiras, R. C. Ashoori, P. Jarillo-Herrero, Correlated insulator behaviour at half-filling in magic-angle graphene superlattices. *Nature* **556**, 80–84 (2018).

[4] A. Chernikov, T. C. Berkelbach, H. M. Hill, A. Rigosi, Y. L. Li, O. B. Aslan, D. R. Reichman, M. S. Hybertsen, T. F. Heinz, Exciton binding energy and nonhydrogenic Rydberg series in monolayer $WS_2$. *Phys. Rev. Lett.* **113**, 076802 (2014).

[5] D. Xiao, G. B. Liu, W. X. Feng, X. D. Xu, W. Yao, Coupled spin and valley physics in monolayers of $MoS_2$ and other group-VI dichalcogenides. *Phys. Rev. Lett.* **108**, 196802 (2012).

[6] J. Kang, S. Tongay, J. Zhou, J. B. Li, J. Q. Wu, Band offsets and heterostructures of two-dimensional semiconductors. *Appl. Phys. Lett.* **102**, 012111 (2013).

[7] P. Rivera, J. R. Schaibley, A. M. Jones, J. S. Ross, S. F. Wu, G. Aivazian, P. Klement, K. Seyler, G. Clark, N. J. Ghimire, J. Q. Yan, D. G. Mandrus, W. Yao, X. D. Xu, Observation of long-lived interlayer excitons in monolayer $MoSe_2$-$WSe_2$ heterostructures. *Nat. Commun.* **6**, 6242 (2015).

[8] P. Rivera, K. L. Seyler, H. Y. Yu, J. R. Schaibley, J. Q. Yan, D. G. Mandrus, W. Yao, X. D. Xu, Valley-polarized exciton dynamics in a 2D semiconductor heterostructure. *Science* **351**, 688–691 (2016).

[9] A. Ciarrocchi, D. Unuchek, A. Avsar, K. Watanabe, T. Taniguchi, A. Kis, Polarization switching and electrical control of interlayer excitons in two-dimensional van der Waals heterostructures. *Nat. Photon.* **13**, 131–136 (2019).

[10] L. A. Jauregui, A. Y. Joe, K. Pistunova, D. S. Wild, A. A. High, Y. Zhou, G. Scuri, K. D. Greve, A. Sushko, C.-H. Yu, T. Taniguchi, K. Watanabe, D. J. Needleman, M. D. Lukin, H. Park, P. Kim, Electrical control of interlayer exciton dynamics in atomically thin heterostructures. *Science* **366**, 870–875 (2019).

[11] H. Y. Yu, Y. Wang, Q. J. Tong, X. D. Xu, W. Yao, Anomalous light cones and valley optical selection rules of interlayer excitons in twisted heterobilayers. *Phys. Rev. Lett.* **115**, 187002 (2015).

[12] P. K. Nayak, Y. Horbatenko, S. Ahn, G. Kim, J. U. Lee, K. Y. Ma, A. R. Jang, H. Lim, D. Kim, S. Ryu, H. Cheong, N. Park, H. S. Shin, Probing evolution of twist-angle-dependent interlayer excitons in $MoSe_2$/$WSe_2$ van der Waals heterostructures. *ACS Nano* **11**, 4041–4050 (2017).

[13] E. M. Alexeev, D. A. Ruiz-Tijerina, M. Danovich, M. J. Hamer, D. J. Terry, P. K. Nayak, S. Ahn, S. Pak, J. Lee, J. I. Sohn, M. R. Molas, M. Koperski, K. Watanabe, T. Taniguchi, K. S. Novoselov, R. V. Gorbachev, H. S. Shin, V. I. Fal'ko, A. I. Tartakovskii, Resonantly hybridized excitons in moire superlattices in van der Waals heterostructures. *Nature* **567**, 81–86 (2019).

[14] D. A. Ruiz-Tijerina, V. I. Fal'ko, Interlayer hybridization and moire superlattice minibands for electrons and excitons in heterobilayers of transition-metal dichalcogenides. *Phys. Rev. B* **99**, 125424 (2019).

[15] H. Y. Yu, G. B. Liu, J. J. Tang, X. D. Xu, W. Yao, Moire excitons: From programmable quantum emitter arrays to spin-orbit-coupled artificial lattices. *Science Advances* **3**, e1701696 (2017).

[16] F. C. Wu, T. Lovorn, A. H. MacDonald, Theory of optical absorption by interlayer excitons in transition metal dichalcogenide heterobilayers. *Phys. Rev. B* **97**, 035306 (2018).

[17] Q. J. Tong, H. Y. Yu, Q. Z. Zhu, Y. Wang, X. D. Xu, A. Yao, Topological mosaics in moiré superlattices of van der Waals heterobilayers. *Nat. Phys.* **13**, 356–362 (2017).

[18] F. C. Wu, T. Lovorn, A. H. MacDonald, Topological exciton bands in moiré Heterojunctions. *Phys. Rev. Lett.* **118**, 147401 (2017).

[19] J. Perczel, J. Borregaard, D. E. Chang, H. Pichler, S. F. Yelin, P. Zoller, M. D. Lukin, Topological quantum optics in two-dimensional atomic arrays. *Phys. Rev. Lett.* **119**, 023603 (2017).

[20] K. L. Seyler, P. Rivera, H. Y. Yu, N. P. Wilson, E. L. Ray, D. G. Mandrus, J. Q. Yan, W. Yao, X. D. Xu, Signatures of moiré-trapped valley excitons in $MoSe_2/WSe_2$ heterobilayers. *Nature* **567**, 66–70 (2019).

[21] M. Brotons-Gisbert, H. Baek, A. Molina-Sánchez, D. Scerri, D. White, K. Watanabe, T. Taniguchi, C. Bonato, B. D. Gerardot, Spin-layer locking of interlayer valley excitons trapped in moiré potentials. https://arxiv.org/abs/1908.03778 (2019).

[22] K. Tran, G. Moody, F. C. Wu, X. B. Lu, J. Choi, K. Kim, A. Rai, D. A. Sanchez, J. M. Quan, A. Singh, J. Embley, A. Zepeda, M. Campbell, T. Autry, T. Taniguchi, K. Watanabe, N. S. Lu, S. K. Banerjee, K. L. Silverman, S. Kim, E. Tutuc, L. Yang, A. H. MacDonald, X. Q. Li, Evidence for moiré excitons in van der Waals heterostructures. *Nature* **567**, 71–75 (2019).

[23] C. H. Jin, E. C. Regan, A. M. Yan, M. I. B. Utama, D. Q. Wang, S. H. Zhao, Y. Qin, S. J. Yang, Z. R. Zheng, S. Y. Shi, K. Watanabe, T. Taniguchi, S. Tongay, A. Zettl, F. Wang, Observation of moire excitons in $WSe_2/WS_2$ heterostructure superlattices. *Nature* **569**, 76–80 (2019).

[24] C. H. Jin, E. C. Regan, D. Q. Wang, M. I. B. Utama, C. S. Yang, J. Cain, Y. Qin, Y. X. Shen, Z. R. Zheng, K. Watanabe, T. Taniguchi, S. Tongay, A. Zettl, F. Wang, Identification of spin, valley and moire quasi-angular momentum of interlayer excitons. *Nat. Phys.* **15**, 1140–1144 (2019).

[25] P. Michler, A. Kiraz, C. Becher, W. V. Schoenfeld, P. M. Petroff, L. D. Zhang, E. Hu, A. Imamoglu, A quantum dot single-photon turnstile device. *Science* **290**, 2282–2285 (2000).

[26] A. Srivastava, M. Sidler, A. V. Allain, D. S. Lembke, A. Kis, A. Imamoglu, Optically active quantum dots in monolayer $WSe_2$. *Nat. Nanotechnol.* **10**, 491–496 (2015).

[27] M. Brotons-Gisbert, A. Branny, S. Kumar, R. Picard, R. Proux, M. Gray, K. S. Burch, K. Watanabe, T. Taniguchi, B. D. Gerardot, Coulomb blockade in an atomically thin quantum dot coupled to a tunable Fermi reservoir. *Nat. Nanotechnol.* **14**, 442–446 (2019).

[28] M. Brotons-Gisbert, J. P. Martinez-Pastor, G. C. Ballesteros, B. D. Gerardot, J. F. Sanchez-Royo, Engineering light emission of two-dimensional materials in both the weak and strong coupling regimes. *Nanophotonics* **7**, 253–267 (2018).

[29] A. T. Hanbicki, H. J. Chuang, M. R. Rosenberger, C. S. Hellberg, S. V. Sivaram, K. M. McCreary, I. I. Mazin, B. T. Jonker, Double indirect interlayer exciton in a $MoSe_2/WSe_2$ van der Waals heterostructure. *ACS Nano* **12**, 4719–4726 (2018).

[30] C. Y. Jiang, W. G. Xu, A. Rasmita, Z. M. Huang, K. Li, Q. H. Xiong, W. B. Gao, Microsecond dark-exciton valley polarization memory in two-dimensional heterostructures. *Nat. Commun.* **9**, 753 (2018).

[31] P. Nagler, M. V. Ballottin, A. A. Mitioglu, F. Mooshammer, N. Paradiso, C. Strunk, R. Huber, A. Chernikov, P. C. M. Christianen, C. Schuller, T. Korn, Giant magnetic splitting inducing near-unity valley polarization in van der Waals heterostructures. *Nat. Commun.* **8**, 1551 (2017).

[32] J. G. Roch, N. Leisgang, G. Froehlicher, P. Makk, K. Watanabe, T. Taniguchi, C. Schonenberger, R. J. Warburton, Quantum-confined stark effect in a $MoS_2$ monolayer van der Waals heterostructure. *Nano Lett.* **18**, 1070–1074 (2018).

[33] K. Kim, S. Larentis, B. Fallahazad, K. Lee, J. M. Xue, D. C. Dillen, C. M. Corbet, E. Tutuc, Band alignment in $WSe_2$-graphene heterostructures. *ACS Nano* **9**, 4527–4532 (2015).

[34] A. Laturia, M. L. Van de Put, W. G. Vandenberghe, Dielectric properties of hexagonal boron nitride and transition metal dichalcogenides: from monolayer to bulk. *NPJ 2D Mater. Appl.* **2**, 6 (2018).

[35] B. Miller, A. Steinhoff, B. Pano, J. Klein, F. Jahnke, A. Holleitner, U. Wurstbauer, Long-lived direct and indirect interlayer excitons in van der Waals heterostructures. *Nano Lett.* **17**, 5229–5237 (2017).

[36] C. D. Zhang, C. P. Chuu, X. B. Ren, M. Y. Li, L. J. Li, C. H. Jin, M. Y. Chou, C. K. Shih, Interlayer couplings, moiré patterns, and 2D electronic

superlattices in $MoS_2/WSe_2$ hetero-bilayers. *Science Advances* **3**, e1601459 (2017).

[37] F. C. Wu, T. Lovorn, E. Tutuc, A. H. MacDonald, Hubbard model physics in transition metal dichalcogenide moiré bands. *Phys. Rev. Lett.* **121**, 026402 (2018).

[38] F. C. Wu, T. Lovorn, E. Tutuc, I. Martin, A. H. MacDonald, Topological insulators in twisted transition metal dichalcogenide homobilayers. *Phys. Rev. Lett.* **122**, 086402 (2019).

[39] Y. Tang, L. Li, T. Li, Y. Xu, S. Liu, K. Barmak, K. Watanabe, T. Taniguchi, A. H. MacDonald, J. Shan, K. F. Mak, $WSe_2/WS_2$ moiré superlattices: a new Hubbard model simulator. https://arxiv.org/abs/1910.08673.

[40] E. C. Regan, D. Wang, C. Jin, M. I. B. Utama, B. Gao, X. Wei, S. Zhao, W. Zhao, K. Yumigeta, M. Blei, J. Carlstroem, K. Watanabe, T. Taniguchi, S. Tongay, M. Crommie, A. Zettl, F. Wang, Optical detection of Mott and generalized Wigner crystal states in $WSe_2/WS_2$ moiré superlattices. https://arxiv.org/abs/1910.09047.

[41] Y. Shimazaki, I. Schwartz, K. Watanabe, T. Taniguchi, M. Kroner, A. Imamoğlu, Moiré superlattice in a $MoSe_2/hBN/MoSe_2$ heterostructure: from coherent coupling of inter- and intra-layer excitons to correlated Mott-like states of electrons. https://arxiv.org/abs/1910.13322.

# Supplementary Materials

### Section S1. Polarization-resolved PL spectroscopy

Fig. S1 shows circular polarization-resolved PL spectra of moiré IXs. For excitation with circular polarization, the excitation energy is set to 1.75 eV for clearer degree of circular polarization. The peaks show co-polarized characteristics.

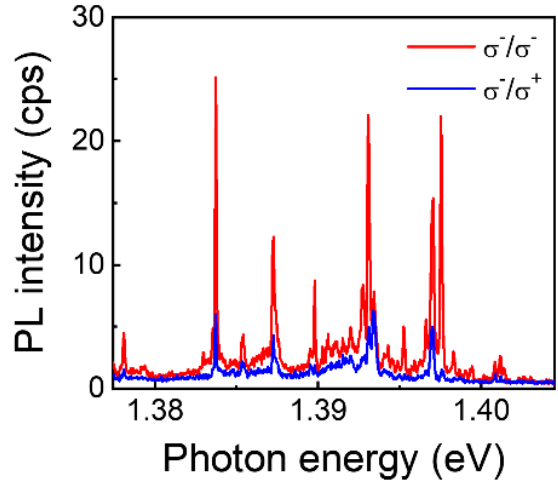

**Fig. S1. Circular polarization of moiré IXs.** PL spectra of moiré IXs with circular polarization of $\sigma^-$ (red) and $\sigma^+$ (blue) under $\sigma^-$ excitation.

Fig. S2 shows PL $\sigma^-$ spectra using $\sigma^-$ excitation as a function of magnetic field. All of peaks shift in the same direction as the magnetic field changes, implying moiré excitons are confined in the same atomic configuration, likely $H_h^h$ (*15, 21*), indicated as blue circle in Fig. 1A.

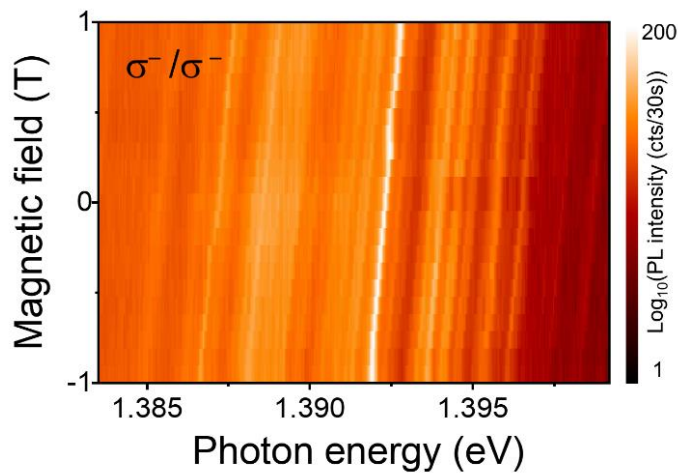

**Fig. S2. Magnetic field dependence of polarization-resolved PL emission.** PL spectra ($\sigma^-$ only) measured at different magnetic field from -1 to 1 T.

## Section S2. Magneto-PL spectroscopy in a second moiré heterostructure sample

Figure S3(a) shows magnetic field dependence of PL emission in a second heterostructure sample. All emission lines display a linear Zeeman splitting with uniform g-factors. As shown in Fig. S3(b), g-factors for peaks at 1.405 and 1.402 eV are very similar, and these values are in a good accordance with previously reported ones for 2*H* stacked heterostructures.

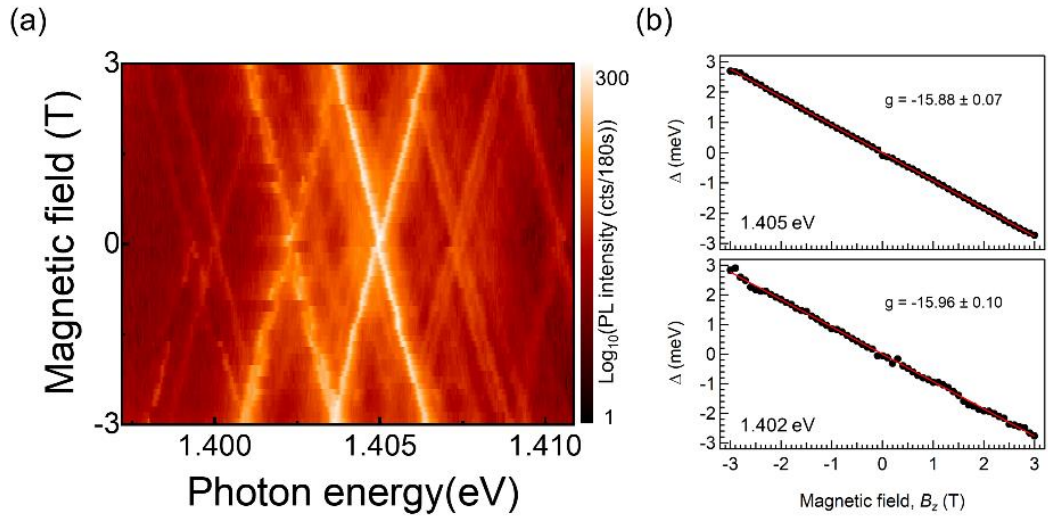

**Fig. S3. Magnetic field dependence of PL emission in Sample 2.** (a) Magnetic field dependence of the moiré IXs as a function of the applied out-of-plane magnetic field. (b) Plots of Zeeman splitting energy versus magnetic field with linear fits for the peaks at 1.405 and 1.402 eV.

## Section S3. Experimental limitation of second-order correlation at zero delay

As discussed in the main manuscript, the experimental value for the normalized second-order photon correlation at zero delay $g^{(2)}(0)$ shown in Fig. 4D is limited by non-filtered emission background. Figure S4(a) shows a high-spectral-resolution PL spectrum of the moiré quantum emitter at 1.401 eV acquired with the same experimental conditions employed for the second-order photon correlation measurements shown in Fig. 4D of the main manuscript prior to spectral filtering. The red and blue shadowed regions represent the estimated background and PL emission of the emitter as obtained from a Lorentzian fit of the PL peak. After spectral filtering with a band-pass filter centered at the emission energy of the emitter (see top panel of Fig. S4(b)). The *SBR* of 6.38 is estimated (see bottom panel of Fig. S4(b)), which limits the experimental determination of $g^{(2)}(0)$ to a minimum a value of $g^{(2)}(0) = 0.25$. In order to estimate the experimental error in the determination of the limitation for $g^{(2)}(0)$, we measured the temporal evolution of the PL of the over 5 minutes (see Fig. S4(c)). As can be seen in this figure, the emission energy of the emitter presents spectral wandering, which results in a time-dependent *SBR* after spectral filtering of the PL spectrum as a consequence of the detuning of the PL signal from the central energy of the filter. Figure S4(d) shows the temporal evolution of the integrated PL intensity of the emitter through the spectral filter over 5 minutes. Unfortunately, this emitter, which was most optimal in terms of signal to noise, exhibited unusually large spectral fluctuations which led to intensity fluctuations after the filter. The dashed line and the yellow shadowed area represent the average value and the standard deviation of the integrated PL intensity, respectively, from which we estimate an experimental error ~0.06 in the determination of the limitation for $g^{(2)}(0)$.

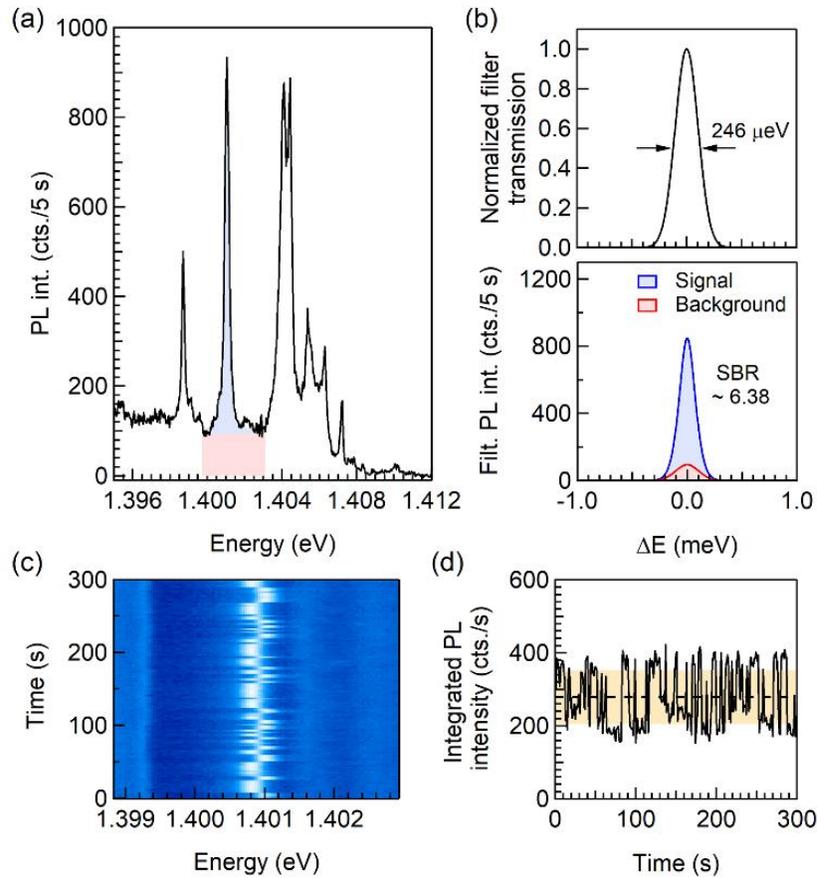

**Fig. S4. Limitation in the determination of $g^{(2)}(0)$.** (a) High-spectral-resolution PL spectrum (prior to spectral filtering) of the moiré quantum emitter acquired with the same experimental conditions employed for the second-order photon correlation measurements shown in Fig. 4D of the main manuscript. The red and blue shadowed regions represent the estimated background and PL emission of the emitter as obtained from a Lorentzian fit of the PL peak. (b) Normalized experimental transmission of the band-pass filter used to spectrally isolate the PL signal of the emitter at 1.401 eV. The bottom panel shows the experimental *SBR* estimated for the emitter after spectral filtering. (c) Temporal evolution of the PL signal of the emitter over 5 minutes. (d) Temporal evolution of the integrated PL intensity of the emitter after spectral filtering over 5 minutes. The dashed line and the yellow shadowed area represent the average value and the standard deviation of the integrated PL intensity, respectively.